
\input phyzzm


\catcode`@=11 

%
\paperstyle   
%
%
\def\MEMO{\letterstyle\FRONTPAGE \letterfrontheadline={\hfil}
    \line{\quad\fourteenrm MEMORANDUM\hfil\twelverm\the\date\quad}
    \medskip \memod@f}

\def\memit@m#1{\smallskip \hangafter=0 \hangindent=1in
      \Textindent{\caps #1}}
\def\memod@f{\xdef\to{\memit@m{To:}}\xdef\from{\memit@m{From:}}%
     \xdef\topic{\memit@m{Topic:}}\xdef\subject{\memit@m{Subject:}}%
     \xdef\rule{\bigskip\hrule height 1pt\bigskip}}
\memod@f
\newskip\lettertopfil
\lettertopfil = 0pt plus 1.5in minus 0pt
\newskip\letterbottomfil
\letterbottomfil = 0pt plus 2.3in minus 0pt
\newskip\spskip \setbox0\hbox{\ } \spskip=-1\wd0
\def\addressee#1{\medskip\rightline{\the\date\hskip 30pt} \bigskip
   \vskip\lettertopfil
   \ialign to\hsize{\strut ##\hfil\tabskip 0pt plus \hsize \cr #1\crcr}
   \medskip\noindent\hskip\spskip}
\newskip\signatureskip       \signatureskip=40pt
\def\signed#1{\par \penalty 9000 \bigskip \dt@pfalse
  \everycr={\noalign{\ifdt@p\vskip\signatureskip\global\dt@pfalse\fi}}
  \setbox0=\vbox{\singlespace \halign{\tabskip 0pt \strut ##\hfil\cr
   \noalign{\global\dt@ptrue}#1\crcr}}
  \line{\hskip 0.5\hsize minus 0.5\hsize \box0\hfil} \medskip }

\def\endletter{\ifnum\pagenumber=1 \vskip\letterbottomfil\supereject
\else \vfil\supereject \fi}
\newbox\letterb@x
\def\lettertext{\par\unvcopy\letterb@x\par}
\def\multiletter{\setbox\letterb@x=\vbox\bgroup
      \everypar{\vrule height 1\baselineskip depth 0pt width 0pt }
      \singlespace \topskip=\baselineskip }
\def\letterend{\par\egroup}
%
%
%
\newskip\frontpageskip
\newtoks\pubtype
\newtoks\Pubnum
\newtoks\pubnum
\newif\ifp@bblock  \p@bblocktrue
\def\PH@SR@V{\doubl@true \baselineskip=24.1pt plus 0.2pt minus 0.1pt
             \parskip= 3pt plus 2pt minus 1pt }
\def\PHYSREV{\paperstyle\PhysRevtrue\PH@SR@V}
\def\titlepage{\FRONTPAGE\paperstyle\ifPhysRev\PH@SR@V\fi
   \ifp@bblock\p@bblock\fi}
\def\nopubblock{\p@bblockfalse}
\def\endpage{\vfil\break}
\frontpageskip=1\medskipamount plus .5fil
\pubtype={ }
\newtoks\publevel
\publevel={Report}   
\Pubnum={}
%
\def\p@bblock{\begingroup \tabskip=\hsize minus \hsize
   \baselineskip=1.5\ht\strutbox \topspace-2\baselineskip
   \halign to\hsize{\strut ##\hfil\tabskip=0pt\crcr
   \the\Pubnum\cr \the\date\cr }\endgroup}
\def\title#1{\vskip\frontpageskip \titlestyle{#1} \vskip\headskip }
\def\author#1{\vskip\frontpageskip\titlestyle{\twelvecp #1}\nobreak}
\def\andauthor{\vskip\frontpageskip\centerline{and}\author}

\def\address#1{\par\kern 5pt\titlestyle{\twelvepoint\it #1}}
\def\andaddress{\par\kern 5pt \centerline{\sl and} \address}


%
\def\abstract{\vskip\frontpageskip\centerline{\fourteenrm ABSTRACT}
              \vskip\headskip }

%
%
%

\def\\{\relax\ifmmode\backslash\else$\backslash$\fi}
\def\globaleqnumbers{\relax\if\equanumber<0\else\global\equanumber=-1\fi}
\def\nextline{\unskip\nobreak\hskip\parfillskip\break}

\def\journal#1&#2(#3){\unskip, \sl #1~\bf #2 \rm (19#3) }

\def\topspace{\hrule height 0pt depth 0pt \vskip}
\def\prop{\mathrel{{\mathchoice{\pr@p\scriptstyle}{\pr@p\scriptstyle}{
                \pr@p\scriptscriptstyle}{\pr@p\scriptscriptstyle} }}}
\def\pr@p#1{\setbox0=\hbox{$\cal #1 \char'103$}
   \hbox{$\cal #1 \char'117$\kern-.4\wd0\box0}}
\def\lsim{\mathrel{\mathpalette\@versim<}}
\def\gsim{\mathrel{\mathpalette\@versim>}}
\def\@versim#1#2{\lower0.2ex\vbox{\baselineskip\z@skip\lineskip\z@skip
  \lineskiplimit\z@\ialign{$\m@th#1\hfil##\hfil$\crcr#2\crcr\sim\crcr}}}
%
%
%
\let\sec@nt=\sec
\def\sec{\relax\ifmmode\let\n@xt=\sec@nt\else\let\n@xt\section\fi\n@xt}
\def\obsolete#1{\message{Macro \string #1 is obsolete.}}
\def\firstsec#1{\obsolete\firstsec \section{#1}}
\def\firstsubsec#1{\obsolete\firstsubsec \subsection{#1}}
\def\thispage#1{\obsolete\thispage \global\pagenumber=#1\frontpagefalse}
\def\thischapter#1{\obsolete\thischapter \global\chapternumber=#1}
\def\nextequation#1{\obsolete\nextequation \global\equanumber=#1
   \ifnum\the\equanumber>0 \global\advance\equanumber by 1 \fi}
\def\BOXITEM{\afterassigment\B@XITEM\setbox0=}
\def\B@XITEM{\par\hangindent\wd0 \noindent\box0 }
%

%
\catcode`@=12 

%
%
\catcode`@=11
%
%

%
\font\fourteentt=cmtt10 scaled\magstep2  
\def\seventeenpoint{\relax
    \textfont0=\seventeenrm         \scriptfont0=\twelverm
    \scriptscriptfont0=\tenrm
     \def\rm{\fam0 \seventeenrm \f@ntkey=0 }\relax
    \textfont1=\seventeeni          \scriptfont1=\twelvei
    \scriptscriptfont1=\teni
     \def\oldstyle{\fam1 \seventeeni\f@ntkey=1 }\relax
    \textfont2=\seventeensy         \scriptfont2=\twelvesy
    \scriptscriptfont2=\tensy
    \textfont3=\seventeenex     \scriptfont3=\seventeenex
    \scriptscriptfont3=\seventeenex
    \def\it{\fam\itfam \seventeenit\f@ntkey=4 }
         \textfont\itfam=\seventeenit
    \def\sl{\fam\slfam \seventeensl\f@ntkey=5 }
         \textfont\slfam=\seventeensl
    \scriptfont\slfam=\twelvesl
    \def\bf{\fam\bffam \seventeenbf\f@ntkey=6 }
         \textfont\bffam=\seventeenbf
    \scriptfont\bffam=\twelvebf  \scriptscriptfont\bffam=\tenbf
    \def\tt{\fam\ttfam \fourteentt \f@ntkey=7 }
         \textfont\ttfam=\fourteentt
    \h@big=11.9\p@{} \h@Big=16.1\p@{} \h@bigg=20.3\p@{} \h@Bigg=24.5\p@{}
    \setbox\strutbox=\hbox{\vrule height 12pt depth 5pt width\z@}
    \samef@nt}
%

%
%
%
\mathchardef\bfalpha  ="090B
\mathchardef\bfbeta   ="090C
\mathchardef\bfgamma  ="090D
\mathchardef\bfdelta  ="090E
\mathchardef\bfepsilon="090F
\mathchardef\bfzeta   ="0910
\mathchardef\bfeta    ="0911
\mathchardef\bftheta  ="0912
\mathchardef\bfiota   ="0913
\mathchardef\bfkappa  ="0914
\mathchardef\bflambda ="0915
\mathchardef\bfmu     ="0916
\mathchardef\bfnu     ="0917
\mathchardef\bfpi     ="0918
\mathchardef\bfrho    ="091A
\mathchardef\bfsigma  ="091B
\mathchardef\bftau    ="091C
\mathchardef\bfupsilon="091D
\mathchardef\bfphi    ="091E
\mathchardef\bfchi    ="091F
\mathchardef\bfpsi    ="0920
\mathchardef\bfomega  ="0921
%
%
\newtoks\heth
\newtoks\Heth
\Pubnum={KUNS \the\pubnum}
\Heth={HE(TH)\the\heth}
\date={\monthname,\ \number\year}
\pubnum={000}
\heth={00/00}
\def\titlepage{\FRONTPAGE\ifPhysRev\PH@SR@V\fi
   \ifp@bblock\p@bblock\fi}
\def\p@bblock{\begingroup \tabskip=\hsize minus \hsize
   \baselineskip=1.5\ht\strutbox \topspace-2\baselineskip
   \halign to\hsize{\strut ##\hfil\tabskip=0pt\crcr
   \the\Pubnum\cr \the\Heth\cr \the\date\cr }\endgroup}
\def\titlestyleb#1{\par\begingroup \interlinepenalty=9999
     \leftskip=0.02\hsize plus 0.23\hsize minus 0.02\hsize
     \rightskip=\leftskip \parfillskip=0pt
     \hyphenpenalty=9000 \exhyphenpenalty=9000
     \tolerance=9999 \pretolerance=9000
     \spaceskip=0.333em \xspaceskip=0.5em
     \iftwelv@\fourteenpoint\else\twelvepoint\fi
   \noindent {\bf #1}\par\endgroup }
\def\title#1{\vskip\frontpageskip \titlestyleb{#1} \vskip\headskip }
%
%

%
%
\paperfootline={\hss\iffrontpage\else\ifp@genum%
                \tenrm --\thinspace\folio\thinspace --\hss\fi\fi}
%
%

%

%
%

%
\def\contr#1#2#3{\vbox{\ialign{##\crcr
          \hskip #2pt\vrule depth 4pt
          \hrulefill\vrule depth 4pt\hskip #3pt
          \crcr\noalign{\kern-1pt\vskip0.125cm\nointerlineskip}
          $\hfil\displaystyle{#1}\hfil$\crcr}}}
\def\leftrightarrowfill{$\m@th\mathord-\mkern-6mu%
  \cleaders\hbox{$\mkern-2mu\mathord-\mkern-2mu$}\hfill
  \mkern-6mu\mathord\leftrightarrow$}
\def\overleftrightarrow#1{\vbox{\ialign{##\crcr
      \leftrightarrowfill\crcr\noalign{\kern-\p@\nointerlineskip}
      $\hfil\displaystyle{#1}\hfil$\crcr}}}
%

%
%

%
\def\rbox#1{\vbox{\hrule height.8pt%
                \hbox{\vrule width.8pt\kern5pt
                \vbox{\kern5pt\hbox{#1}\kern5pt}\kern5pt
                \vrule width.8pt}
                \hrule height.8pt}}
%
%
\def\sqr#1#2{{\vcenter{\hrule height.#2pt
      \hbox{\vrule width.#2pt height#1pt \kern#1pt
          \vrule width.#2pt}
      \hrule height.#2pt}}}
\def\overbar#1{\vbox{\ialign{##\crcr
          \hskip 1.5pt\hrulefill\hskip 1.1pt
          \crcr\noalign{\kern-1pt\vskip0.125cm\nointerlineskip}
          $\hfil\displaystyle{#1}\hfil$\crcr}}}
%
%
%

%
%

%
%
\mathchardef\Lag="724C
%
%

%

%
\def\addeqno{\ifnum\equanumber<0 \global\advance\equanumber by -1
    \else \global\advance\equanumber by 1\fi}
\def\undereq#1{\mathop{\vtop{\ialign{##\crcr
      $\hfil\displaystyle{#1}\hfil$
      \crcr\noalign{\kern3\p@\nointerlineskip}
      \crcr\noalign{\kern3\p@}}}}\limits}
\def\overeq#1{\mathop{\vbox{\ialign{##\crcr\noalign{\kern3\p@}
      \crcr\noalign{\kern3\p@\nointerlineskip}
      $\hfil\displaystyle{#1}\hfil$\crcr}}}\limits}
%

%
%

%
%
\def\journal#1&#2(#3){\unskip, {\sl #1}{\bf #2}(19#3)}
\def\andjournal#1&#2(#3){{\sl #1}{\bf #2}(19#3)}
\def\andvol&#1(#2){{\bf #1}(19#2)}

\def\NP{Nucl. Phys. }
\def\PR{Phys. Rev. }
\def\PRL{Phys. Rev. Lett. }
\def\PL{Phys. Lett. }

\def\NC{Nuovo Cimento }
%
%

%
%

%
%

%
\catcode`@=12
%
%

%

%

\catcode`@=11

\newtoks\heth
\newtoks\Heth
\newtoks\rims
\Pubnum={KUNS \the\pubnum}
\Heth={HE(TH)\the\heth}
\rims={RIMS-778}
\date={Revised September,\ \number\year}
\def\p@bblock{\begingroup \tabskip=\hsize minus \hsize
   \baselineskip=1.5\ht\strutbox \topspace-2\baselineskip
   \halign to\hsize{\strut ##\hfil\tabskip=0pt\crcr
   \the\Pubnum\cr \the\Heth\cr
    \the\rims\cr \the\date\cr }\endgroup}

\catcode`@=12


\def\MPL{Mod.~Phys.~Lett. }

\def\symbol#1/#2/#3/#4/#5/#6/{
     \left( \matrix{#1 & #2 & #3 \cr #4 & #5 & #6 \cr}  \right) }
\def\rwsymbol#1/#2/#3/#4/#5/#6/{
     \left\{  \matrix{#1 & #2 & #3 \cr #4 & #5 & #6 \cr}
\right\}^{RW} }
\def\wsymbol#1/#2/#3/#4/#5/#6/{
     \left\{  \matrix{#1 & #2 & #3 \cr #4 & #5 & #6 \cr}  \right\}}
\def\qsymbol#1/#2/#3/#4/#5/#6/{
     \left\{  \matrix{#1 & #2 & #3 \cr #4 & #5 & #6 \cr}  \right\}_q}
\def\qwsymbol#1/#2/#3/#4/#5/#6/{
     \left|  \matrix{#1 & #2 & #3 \cr #4 & #5 & #6 \cr}  \right|_q}

\def\CMP{Commun.~Math.~Phys. }

\pubnum={1088}          
\heth={91/13}           


\REF\ADF{J.~Ambj\o rn, B.~Durhuus and J.~Fr\"olich \journal \NP &B257
[FS14] (85) 433; \nextline
F.~David \journal \NP &B257 [FS14] (85) 45; \nextline
V.A.~Kazakov, I.K.~Kostov and A.A.~Migdal \journal \PL &157B (85) 295.}

\REF\KPZ{V.G.~Knizhnik, A.M.~Polyakov and A.B.~Zamolodchikov
\journal \MPL &A3 (88) 819.}
\REF\DDK{F.~David \journal \MPL &A3 (88) 1651; \nextline
J.~Distler and H.~Kawai \journal \NP &B321 (89) 509.}

\REF\Kazakov{V.A.~Kazakov \journal \MPL &A4 (89) 2125.}
\REF\GM{
  M.R.~Douglas and S.H.~Shenker \journal \NP &B335 (90) 635;
  \nextline
  D.J.~Gross and A.A.~Migdal \journal \PRL &64 (90) 127;
  \nextline
  E.~Brezin and V.A.~Kazakov \journal \PL &236B (90) 144.}
\REF\Regge{G.~Ponzano and T.~Regge, in {\it Spectroscopic and Group
Theoretical Methods in Physics, }ed. F.~Bloch (North-Holland
, Amsterdam, 1968). }
\REF\reggeone{T.~Regge \journal \NC &19 (61) 558.}
\REF\qg{M.~Jimbo, {\sl Lett. Math. Phys.} {\bf 10} (1985) 63;
\nextline
V.G.~Drinfeld, ``{\it Quantum Groups},'' in {\it Proc.~Intl.~Congress of
Mathematicians}, Berkeley, California, (1986) p.798.}
\REF\Turaevone{V.G.~Turaev and O.Y.~Viro,
``{\it State Sum Invariants of 3-Manifolds
and Quantum $6j$-Symbols,}'' preprint (1990). }
\REF\alexander{J.W.~Alexander, {\sl Ann. Math.} {\bf 31}
(1930) 292.}
\REF\Alvarez{A.~Tsuchiya and Y.~Kanie, in {\it Conformal Field Theory
and Solvable Lattice Models},
Advanced Studies in Pure Mathematics, {\bf 16} (1988) 297;
\nextline
T.~Kohno, in ``{\it Quantized Universal Enveloping Algebras and
Monodromy of Braid Groups,}'' Nagoya preprint (1988);
\nextline
L.~Alvarez-Gaum\'e, C.~Gomez and G.~Sierra
\journal \PL &220B (89) 142;
\nextline
G.~Moore and N.~Reshetikhin \journal \NP &B328 (89) 557.}
\REF\moore{G.~Moore and N.~Seiberg \journal \PL &212B (88) 451;
\nextline
\andjournal \CMP &123 (89) 77.}
\REF\cstheory{E.~Witten \journal \CMP &121 (89) 351.}
\REF\dewitt{B.S.~DeWitt \journal \PR &160 (67) 1113; \nextline
J.A.~Wheeler, in {\it Batelle Recontres}, eds. C.M.~DeWitt and
J.A.~Wheeler (W.A.~Benjamin, inc., New York, 1968).}
\REF\cgcoefficient{See for example, D.A.~Varshalovich, A.N.~Moskalev
and V.K.~Khersonskii, {\it Quantum Theory of Angular Momentum}
(World Scientific, Singapore, 1988).}
\REF\network{B.V.~Boulatov, V.A.~Kazakov, I.K.~Kostov
and A.A.~Migdal \journal \NP &B275 [FS17] (86) 641.}
\REF\witten{E.~Witten \journal \NP &B311 (88/89) 46.}
\REF\Turaevtwo{V.G.~Turaev, ``{\it Quantum Invariants of 3-Manifolds and
a Glimpse of \nextline
Shadow Topology,}'' hand-written manuscript (1991).}
\REF\bftheory{I.~Oda and S.~Yahikozawa, preprint IC-90-44 (1990).}
\titlepage

\title{Discrete and Continuum Approaches
  \break to Three-Dimensional Quantum Gravity}

\author{Hirosi Ooguri\foot{e-mail addresses : ooguri@jpnrifp.bitnet,
 and  ooguri::kekvax }}

\address{Research Institute for Mathematical Sciences \break
         Kyoto University,
         Kyoto 606, Japan}

\andauthor{Naoki Sasakura\foot{e-mail address :
sasakura@jpnrifp.bitnet}}

\address{Department of Physics, Kyoto University \break
          Kyoto 606, Japan}

\abstract{It is shown that, in the three-dimensional lattice
gravity defined by Ponzano and Regge, the space of physical
states is isomorphic to the space of gauge-invariant functions on
the moduli
space of flat $SU(2)$ connections over a two-dimensional surface,
which gives physical states in the $ISO(3)$
Chern-Simons gauge theory. To prove this, we employ the
$q$-analogue of this model defined by Turaev and Viro as a
regularization to sum over states. A recent work by Turaev
suggests that the $q$-analogue model itself may be related
to an Euclidean gravity with a cosmological constant proportional
to $1/k^2$, where $q=e^{2\pi i/(k+2)}$.}
\endpage

\sequentialequations

In the two-dimensional quantum gravity, the connection between
the dis-crete\refmark{\ADF}
and the continuum\refmark{\KPZ,\DDK} approaches has been much explored
recently\refmark{\Kazakov,\GM}
and this facilitated our understanding on various
aspects of the theory. In this letter, we point out that
an example also exists in three dimensions where one can compare
these two approaches.

The lattice gravity we consider here is the one originally
defined by Ponzano and Regge.\refmark{\Regge}
Consider a simplicial decomposition of a three-dimensional manifold $M$.
Each three-simplex (tetrahedron) has four faces and six edges.
To each edge,
we assign a half-integral number $j$ ($j=0,{1 \over 2},1,...$) and
regard
it as a ``discretized length'' of the edge.
In this way,
each tetrahedron is colored by
an ordered set of six numbers $(j_1,...,j_6)$, where boundaries of four
faces
of the tetrahedron are colored as $(j_1,j_2,j_3)$, $(j_3,j_4,j_5)$,
$(j_5,j_6,j_1)$ and $(j_6,j_4,j_2)$ respectively.
For the tetrahedron to be realized in the three-dimensional space,
the edge-lengths $j_i$'s must
satisfy the triangle inequalities
(i.e. $|j_1-j_2|\leq j_3 \leq j_1+j_2$ when edges colored as
$j_1$, $j_2$ and $j_3$ meet at a face of the tetrahedron).
This reminds us of a decomposition rule of a tensor product of
$SU(2)$ representations.
Thus it is tempting to
regard the edge-lengths $j_i$'s as highest weights of
$SU(2)$ and introduce
the Racah-Wigner $6j$-symbol
$$ \wsymbol j_1/j_2/j_3/j_4/j_5/j_6/.$$
Since the $6j$-symbol has the
tetrahedral symmetry, we can associate it
to
the colored tetrahedron without ambiguity.

Ponzano and Regge made a remarkable observation that,
when $j$'s are large, the $6j$-symbol is approximated as
$$
\eqalignno{&\exp (-i\pi \sum_i j_i)
   \wsymbol j_1/j_2/j_3/j_4/j_5/j_6/ \sim {1 \over \sqrt{48 \pi V}}
       \left( (-1)^{\sum_i 2 j_i}
                         e^{+i( S_{Regge} - \pi/4)}
         + e^{-i(S_{Regge} - \pi/4)} \right), \cr
&&\eqnalign{\asymptotic} \cr}
$$
where $S_{Regge}$ is the Regge action\refmark{\reggeone}
for the single tetrahedron
$$
  S_{Regge} = \sum_{i=1}^6 (\pi - \theta_i) j_i
$$
with $\theta_i$ being the angle between outer normals of
the two faces belonging to the {$i$-th} edge colored as $j_i$, and
$V$ is the volume of the tetrahedron.
This Regge action gives a discretized version of the Euclidean
Einstein-Hilbert action $\int d^3 x \sqrt{g} R$ assuming that
all the faces of the tetrahedron are flat and a curvature is
concentrated only on their edges.
Now  let us take a product of $6j$-symbols over all the tetrahedra
in $M$. Among various interference terms of $e^{+iS_{Regge}}$
and $e^{-iS_{Regge}}$, the product contains $e^{i \sum S_{Regge}}$
where the sum in the exponent is over all the tetrahedra.
Thus the sum of the product
of the $6j$-symbols over all possible colorings
may approximate the three-dimensional Euclidean
gravity.

There are obvious problems to be overcome before taking this
approach seriously.
First it is not clear why one can ignore the interference
terms and concentrate only on the product of the positive frequency part
$e^{iS_{Regge}}$\foot{One may also worry about the sign factor
$(-1)^{\sum_i 2 j_i}$ in front of $e^{iS_{Regge}}$ in \asymptotic .
This will be taken care of if we restrict $j_i$ to be integral.
With this restriction, most of the following discussions will go
through.}.
Also one needs to know whether there exists
a nice continuum limit of the theory. Since the $6j$-symbol approximates
the Regge action at large values of $j_i$'s, the continuum limit must
be such that the sum over the coloring of the tetrahedra
is dominated by these $j_i$'s.

Actually the sum is divergent at large $j_i$'s. One might hope then,
with an appropriate choice of regularization,
only large values of $j_i$'s become
relevant in the summation and the $6j$-symbol is well-approximated by
the Regge action.
The main purpose of this note is to point out that indeed there exists
a regularization such that the sum over coloring of tetrahedra
described in the above becomes
identical to the continuum functional integral
$$
  Z = \int  \left[  de,  d\omega   \right] \exp( i \int e \wedge R ),
\eqn\iso
$$
where $e$ is a dreibein, $\omega$ is a spin connection on $M$,
and $R$ is a curvature two-form computed from $\omega$
as $R= d\omega + \omega \wedge \omega$.
We will examine physical states
in the Ponzano-Regge model and show that they can be identified with
the ones in the continuum field theory \iso .
There is a disturbing factor of $i$  in front
of the Einstein-Hilbert action $S= \int e \wedge R$
in the exponent of the integrand. We shall discuss on this issue
at the end of this letter.

Now let us discuss the regularization of the sum over coloring.
The method adopted by Ponzano and Regge in their original paper
was  simply  to cut-off the sum by $j_i \leq L$ and rescale it
by multiplying a factor ${1 \over \Lambda(L)}$ per each vertex.
$$\eqalign{
  Z(L) = & \sum_{colorings \atop (j_e \leq L)}
   ~~~ \prod_{vertices} {1 \over \Lambda(L)}~
        \prod_{e:edges} (-1)^{2j_e}(2j_e+1)  \cr &
    ~~~~~~~~  \prod_{t:tetrahedra}
 \exp (-\pi i\sum_i j_i(t))\wsymbol
j_1(t)/j_2(t)/j_3(t)/j_4(t)/j_5(t)/j_6(t)/ . \cr}
\eqn\ponzano
$$
Here $e$ and $t$ run over all the edges and the tetrahedra in $M$,
$j_e$ is a coloring on $e$, and ($j_1(t),...,j_6(t)$)
gives colorings on edges
belonging to $t$.
The rescaling factor $\Lambda(L)$ is given by
$$
\Lambda(L) = {1 \over 2l+1} \sum_{0\leq i,j \leq L,~ |i-j| \leq l
\leq i+j
                                 \atop
                                   i+j+l=0~mod~{\cal Z}}
           (2i+1)(2j+1).
\eqn\rescaling
$$
The right-hand-side in the above does not depend on $l$
as far as $l$ is sufficiently small compared to $L$.
In the limit $L \rightarrow \infty$, the factor $\Lambda(L)$
diverges as $\Lambda(L) \sim 4 L^3 /3 $.

In the light of recent advances in quantum group
technology\refmark{\qg}, however,
it may appear more natural to regularize
the summation by considering
the $q$-analogue of the model.
In the quantum group $U_q(SU(2))$
associated to $q=e^{2\pi i/(k+2)}$ with $k$ being an integer,
representations with $j \leq k/2$ enjoy special status.
Thus
the cut-off of the edge-lengths $j$ will be built in as
an intrinsic property in the $q$-analogue of the Ponzano-Regge model.
Indeed such a model was considered by Turaev and Viro.\refmark{
\Turaevone}
The partition function of their model is obtained by
taking the $q$-analogue of \ponzano\ as\foot{If the
manifold $M$ has some boundaries, we multiply ${1 \over
\sqrt{\Lambda_q}}$
per each vertex on the boundaries and $e^{i \pi j_e}\sqrt{[2j_e +1]_q}$
per each edge in the summand.}
$$
\eqalign{
 Z_k =& \sum_{colorings \atop (j_e \leq k/2)}
 ~~\prod_{vertices} {1 \over \Lambda_q}~
  \prod_{e:edges} (-1)^{2j_e} [ 2j_e+1 ]_q
        \cr &
     ~~~~~\prod_{t:tetrahedra}
     \exp (-\pi i \sum_i j_i(t))
   \qsymbol
j_1(t)/j_2(t)/j_3(t)/j_4(t)/j_5(t)/j_6(t)/ .
    \cr}
\eqn\turaev
$$
Here $\qsymbol j_1/j_2/j_3/j_4/j_5/j_6/$ is the $6j$-symbol of
the quantum group $U_q(SU(2))$ and $[ 2j+1 ]_q$ denotes the $q$-analogue
of the integer $(2j+1)$,
$$
   [ 2j+1 ]_q = { q^{(2j+1) / 2} - q^{-(2j+1)/2}
                 \over q^{1/2} - q^{-1/2} } .
$$
The factor $\Lambda_q$
is defined by generalizing \rescaling\ as
$$
\Lambda_q = {1 \over [2l+1]_q} \sum_{0\leq i,j \leq k/2,~i+j+l \leq k
         \atop
      |i-j| \leq l \leq i+j,~~ i+j+l=0~mod~{\cal Z}}
           [2i+1]_q [2j+1]_q .
\eqn\qscale
$$
$\Lambda_q$ thus defined is independent of $l$.
In fact, one can perform the summation and obtain $\Lambda_q
= -2(k+2)/(q^{1/2}-q^{-1/2})^2$.

The remarkable property of this model is that
the value of
$Z_k$ is independent of a choice of the simplicial decomposition of
the three-dimensional
manifold $M$. It had already been suggested by
Ponzano and Regge that $Z(L \rightarrow \infty)$ is invariant under
a refinement of the decomposition;
one can decompose a single tetrahedron into four
smaller tetrahedra by adding an extra-vertex in the middle of the
original tetrahedron and $Z(\infty)$ does not change its value
under this operation. Turaev and Viro have gone further and shown
that $Z_k$ is invariant under an arbitrary Alexander
transformation of simplices, which
includes the refinement of a single tetrahedron as a special case.
It is known that one can relate any two triangulations
of $M$ by a sequence of Alexander
transformations.\refmark{\alexander} Thus
$Z_k$ depends only on the topology of $M$.

Although there are infinite number of Alexander transformations
in three dimensions,
they are represented as compositions of three local moves.
The partition function $Z_k$ is preserved under those three moves
provided
$$
\sum_j (-1)^{2(j+j_4)} [2j+1]_q  [2j_4+1]_q
\qwsymbol j_2/j_1/j/j_3/j_5/j_4/
                 \qwsymbol j_3/j_1/j_6/j_2/j_5/j/ = \delta _{j_4j_6}
\ ,
\eqn\condone
$$
$$
\eqalign{ &\sum_j (-1)^{2j} [2j+1]_q
\qwsymbol j_2/a/j/j_1/c/b/ \qwsymbol j_3/j/e/j_1/f/c/
\qwsymbol j_3/j_2/j_{23}/a/e/j/ \cr &~~~~~~~~~~~~~~
{}~~~~~=
\qwsymbol j_{23}/a/e/j_1/f/b/
\qwsymbol j_3/j_2/j_{23}/b/f/c/ \cr}
\eqn\condtwo
$$
and \qscale\
are satisfied. Here
$$
 \qwsymbol j_1/j_2/j_3/j_4/j_5/j_6/
   =  \exp (-\pi i \sum_i j_i) \qsymbol
j_1/j_2/j_3/j_4/j_5/j_6/.
$$
These in fact are well-known formulae in conformal field theory.
It has been pointed out by several groups\refmark{\Alvarez,\moore} that
the quantum $6j$-symbol is equal to
the fusion matrix in the $SU(2)$ Wess-Zumino-Witten (WZW)
model
upto some phase factor. The equation \condtwo\ can then be
regarded as
the pentagon identity in the modular tensor category
of Moore and Seiberg\refmark{\moore},
and \condone\ is the unitarity
relation\foot{Thus one could define a large class of
three-dimensional lattice model of this type
associated to various conformal field theories.}.

With this connection to the WZW model, one might suspect
that the lattice model
of Turaev and Viro is related to the $SU(2)$ Chern-Simons (CS)
gauge theory.\refmark{\cstheory}
To see if this is the case, it is useful to introduce a finite
dimensional vector space $H^{(k)}(\Sigma)$, defined by
Turaev and Viro,  associated to a
two-dimensional closed topological surface $\Sigma$.
Let us fix a triangulation $t$ of $\Sigma$ and consider
a vector space $C^{(k)}(\Sigma,t)$ which is freely generated
by all the possible colorings of $t$ over $\bf{C}$.
Now we can define a linear map $Q_{t\rightarrow t}$ from
$C^{(k)}(\Sigma,t)$ into
itself as follows. Let us take a three-dimensional manifold
$M$ to be of the topology $\Sigma \times [0,1]$. $M$ has two
boundaries and both of them are isomorphic to $\Sigma$.
We then fix a simplicial decomposition of $M$ in such a way
that, at
the boundaries of $M$, it agrees with the triangulation $t$
of $\Sigma$. With this preparation, one can compute
a partition function $Z_k(c_1,c_2)$ as in \turaev,
where we fix colorings $c_1$ and $c_2$
on the boundaries
$\Sigma \times \{ 0  \}$ and $\Sigma \times \{ 1  \}$
of $M$.
Now a linear map $Q_{t\rightarrow t}$ is defined as
$$
  Q_{t\rightarrow t}
  : ~ c ~~~\rightarrow \sum_{c' \in C^{(k)}(\Sigma,t)}
             Z_k(c,c') ~~ c'
$$
Due to the invariance of $Z_k$ under the Alexander transformations,
the definition of the map $Q_{t\rightarrow t}$ is independent of a
choice
of the simplicial decomposition  in the interior of $M$.
As a corollary of this, one can show
$(Q_{t\rightarrow t})^2 = Q_{t\rightarrow t}$, namely
$Q_{t \rightarrow t}$ is a projection operator.
Thus an eigenvalue of $Q_{t \rightarrow t}$ is either $0$ or
$1$. We pick a subspace
with an eigenvalue $1$ and call it $H^{(k)}(\Sigma,t)$.
In the Hamiltonian picture of the model,
$Q_{t\rightarrow t}$ may be viewed as a time-evolution operator
associated to the topology $\Sigma \times [0,1]$. The condition
$Q_{t\rightarrow t}=1$ should then correspond to the Hamiltonian
constraint
(or the Wheeler-DeWitt equation\refmark{\dewitt})
in the quantum gravity.

So far, we have considered $H^{(k)}(\Sigma,t)$ with respect to the fixed
triangulation $t$ of $\Sigma$.
One can show that two vector spaces
$H^{(k)}(\Sigma,t_1)$ and $H^{(k)}(\Sigma,t_2)$
associated to different triangulations
$t_1$ and $t_2$
are isomorphic if they are associated to the same topological
surface $\Sigma$.
To show this, we consider again the manifold $M = \Sigma \times
[0,1]$, but this time $\Sigma \times \{ 0 \}$ and
$\Sigma \times \{1 \}$ have different triangulations
$t_1$ and $t_2$. By generalizing the construction of
$Q_{t\rightarrow t}$
in the above, we can define an
operator $Q_{t_1 \rightarrow t_2}$
which maps $C^{(k)}(\Sigma,t_1)$ into $C^{(k)}(\Sigma,t_2)$.
Since $Q_{t_1 \rightarrow t_2} \circ
Q_{t_2 \rightarrow t_1}$ is
equal to the map $Q_{t_1 \rightarrow
t_1}$  which acts as an identity operator on $H^{(k)}
(\Sigma,t_1)$,
the restriction of $Q_{t_1\rightarrow t_2}$ on
$H^{(k)}(\Sigma,t_1)$ defines
an isomorphism between $H^{(k)}(\Sigma,t_1)$ and $H^{(k)}(\Sigma,t_2)$.
Thus we may drop $t$ in $H^{(k)}(\Sigma,t)$ and call it a space of
``physical states'' in the Turaev-Viro model.

Now we are in a position to establish a connection between
the lattice model of Ponzano and Regge and the continuum field
theory given by \iso .
First let us show that the space of physical states
in the Ponzano-Regge model $H^{(k=\infty)}(\Sigma)$
is isomorphic to ${\cal{F}}(flat)$,
the space of gauge-invariant functions over the moduli space of flat
$SU(2)$ connections.
To study the structure of ${\cal{F}}(flat)$,
it is easier to start with
${\cal{F}}$ consisting of gauge-invariant
functions on the space of all
$SU(2)$ connections.  An element of ${\cal{F}}$
is constructed from Wilson-line operators $U_j(x,y)$ ($x,y \in \Sigma$,
$j=0,{1\over2},1,...$),
$$
   U_j(x,y) = P \exp( \int_{x}^{y} A^a t^a_j ),
$$
where $P\exp$ denotes the path ordered exponential and $t^a_j$
($a=1,2,3$)
is the spin-$j$ generator of $SU(2)$. Under a gauge transformation,
$A \rightarrow \Omega^{-1} A \Omega + \Omega^{-1} d \Omega$,
the Wilson-line operator behaves as $U(x,y) \rightarrow
\Omega(x)^{-1} U(x,y) \Omega(y)$.
Now consider their tensor product
$
     \otimes_i U_{j_i} (x_i,y_i)
$.
To make this gauge-invariant, we need to contract group indices
of $U_i$'s so that the gauge factor $\Omega$ cancels out.
Invariant tensors we can use to contract the indices are
the Clebsch-Gordan (CG) coefficient $\langle
j_1 j_2 m_1 m_2 | j_3 m_3 \rangle$ and the metric $g^{(j)}_{m m'}
= \sqrt{2j+1} \langle j j m m' | 0,0 \rangle
= (-1)^{j-m} \delta_{m+m',0} $.
The gauge-invariant function
constructed this way corresponds to a colored trivalent graph $Y$
on $\Sigma$, where a contour from $x$ to $y$ in $Y$
corresponds to a Wilson-line $U(x,y)$ and
three-point vertices in $Y$ represent the CG-coefficients
\foot{Subtlety arises when
there are two Wilson-lines intersecting with each other.
In such a case, we cut the Wilson-lines at the intersecting point and
use the identity $g^{(j_1)}_{m_1n_1} g^{(j_2)}_{m_2n_2}
=\sum_{j,m,n}g^{(j)}_{mn}
\langle j_1 j_2 m_1 m_2 | j m \rangle
\langle j_1 j_2 n_1 n_2 | j n \rangle$ to replace the intersection by
two vertices and an infinitesimal
Wilson-line connecting them.}.
Thus, to each colored trivalent graph $Y$,
we can associate a function $\Psi_Y \in {\cal{F}}$.
In general, a gauge-invariant function of $A$ is a linear
combination of such $\Psi_Y$'s. Thus ${\cal{F}}$ is isomorphic to
a vector space $\tilde{C}(\Sigma)$ which is freely
generated by colored trivalent graphs on $\Sigma$.
The isomorphism is defined as $\sum_i a_i Y_i \in \tilde{C}(\Sigma)
\rightarrow \sum_i a_i \Psi_{Y_i} \in {\cal{F}}$, where $Y_i$'s
are colored trivalent graphs.

The function $\Psi_Y(A)$ can be regarded as an element of
${\cal{F}}(flat)$
by simply restricting its domain to the space of
flat connections. The map  $Y \rightarrow \Psi_Y \in {\cal{F}}(flat)$
however, is not injective (i.e. two
different graphs $Y$ and $Y'$ may give the same
function $\Psi_Y=\Psi_{Y'}$ upon the restriction).
In general, if two graphs $Y$ and $Y'$ are homotopic, the
corresponding functions $\Psi_Y$ and $\Psi_{Y'}$  have the
same value on flat connections.
To each colored trivalent graph $Y$, one can associate a colored
triangulation as a dual graph.
Especially if two graphs $Y$ and $Y'$ are homotopically inequivalent,
they correspond to distinct
colored triangulations.
The isomorphism between $\tilde{C}(\Sigma)$ and ${\cal{F}}$
then induces a homomorphism $\varphi: C(\Sigma)
\rightarrow {\cal{F}}(flat)$, where $C(\Sigma)=\oplus_t
C^{(k=\infty)}(\Sigma,t)$ is a vector space freely generated by colored
triangulations. Thus in order to study the structure
of ${\cal{F}}(flat)$, we would like to identify the kernel of $\varphi$.

Since $\Psi_Y$ on a flat connection is invariant under
homotopy move of $Y$,
we may pick any open Wilson-line in $Y$ and make its length
to be arbitrary small without changing the value of $\Psi_Y$.
The corresponding Wilson operator can then be
replaced by an identity, and group indices of CG-coefficients
at two end-points of the Wilson-line are simply summed
over. Now there is a formula which relates two different ways of
summing CG-coefficients\refmark{\cgcoefficient},
$$
\eqalign{ &
   \sum_{m_6} \langle j_2 j_4 m_2 m_4 | j_6 m_6 \rangle
              \langle j_1 j_6 m_1 m_6 | j_5 m_5 \rangle \cr &=
     \sum_{j_3}
         e^{\pi i (j_3+j_6)}
          \sqrt{(2j_3+1)(2j_6+1)} e^{-\pi i \sum_i j_i}
     \wsymbol j_1/j_2/j_3/j_4/j_5/j_6/ \cr
         &~~~~~~~~~~~~~~~~~~~~
        \sum_{m_3} \langle j_1 j_2 m_1 m_2 |  j_3 m_3 \rangle
                 \langle j_3 j_4 m_3 m_4 | j_5 m_5 \rangle. \cr }
\eqn\id
$$
We can use this to obtain the following relation.
$$
  \Psi_{\tilde{Y}_{j_6}} =  \sum_{j_3} e^{\pi i (j_3+j_6)}
         \sqrt{(2j_3+1)(2j_6+1)}  e^{-\pi i
\sum_i j_i} \wsymbol j_1/j_2/j_3/j_4/j_5/j_6/ \Psi_{Y_{j_3}},
\eqn\move
$$
where the graph $Y_{j_3}$ contains a Wilson-line of spin-$j_3$
connecting lines with
$j_1$ and $j_2$ to lines with $j_4$ and
$j_5$ at the two end-points, while $\tilde{Y}_{j_6}$ is obtained
by replacing this Wilson-line in $Y_{j_3}$ by its dual line of
spin-$j_6$
connecting $j_1$ and $j_5$ to $j_2$ and $j_4$.

If a graph contains a contractible loop with several external lines,
by repeatedly using \id ,
the loop can be recombined into a tree with a one-loop tadpole.
The tadpole can be made arbitrarily small, and
the infinitesimal tadpole can be removed by using
$$
   \sum_{mm'} g_{mm'}^{(j)} \langle j j m m' | J M \rangle
          = \sqrt{2j+1} \delta_{J,0} \delta_{M,0}.
\eqn\idtwo
$$
For example, if $Y$
contains a loop with three external-lines $j_1$, $j_2$ and
$j_3$ attached, we can shrink the loop to obtain another
graph $Y'$ where
the three lines meet at one point. The corresponding functions
$\Psi_Y$
and $\Psi_{Y'}$ are related as
$$
  \Psi_Y
    =  \sqrt{(2l_{12}+1)(2l_{23}+1)
             (2l_{31}+1)} e^{- \pi i \sum_{i=1}^3 j_i}
        \wsymbol j_1/j_2/j_3/l_{23}/l_{31}/l_{12}/ \Psi_{Y'},
\eqn\movetwo
$$
where $l_{ij}$ is the color of the segment
of the loop in $Y$ connecting $j_i$ and
$j_j$. Using a variation of the analysis in Appendix D of [\network ],
one can show that
all other relations among $\Psi_Y$'s in ${\cal{F}}(flat)$
are generated from \move\ and \movetwo .

{}From \move\ and \movetwo , one sees that if $c \in
C^{(k=\infty)}(\Sigma,t)$ and $c' \in C^{(k=\infty)}(\Sigma,t')$
are related as
$c' = Q_{t\rightarrow t'} c$, they are mapped into the same function
in ${\cal{F}}(flat)$ by $\varphi$.
For example, \movetwo\ is realized as a process of
attaching a tetrahedron on a single triangle in $t$,
while \move\ is an operation to recombine two
neighboring triangles on $\Sigma$ into ones in a dual position.
Since the map $Q_{t \rightarrow t'}$ can be constructed from these
two moves,
the kernel of the homomorphism $\varphi$ is characterized
by the relations given by $Q_{t\rightarrow t'}$ on $C(\Sigma)$,
namely $C(\Sigma)/ker(\varphi) \simeq H^{(k=\infty)}(\Sigma)$.
By the homomorphism theorem, we obtain ${\cal{F}}(flat) \simeq
H^{(k=\infty)}(\Sigma)$.

Now that we have found $H^{(k = \infty)}(\Sigma)
\simeq {\cal{F}}(flat)$,
we would like to connect it to the physical Hilbert space of
the continuum field theory \iso . Following the observation
by Witten\refmark{\witten}
in the case of the Lorentzian gravity, we regard
$\int e \wedge R$ as the CS-action whose
gauge group is $ISO(3)$. In the Hamiltonian formulation of
the $ISO(3)$ CS-theory, the timelike components
of $e$ and $\omega$ do not have their canonical conjugate variables,
while the spacelike components
$e_i$ and $\omega_i$ ($i=1,2$) are conjugate to
each other. The variable $e_0$ imposes a constraint that
the $SU(2)$ connection $\omega_i$ on $\Sigma$ is
flat\foot{As we mentioned before, we may either include
half-integral $j$'s in the summation in (3) and (4) or
restrict $j$'s to be integral. In the latter case,
$\omega_i$ should be viewed as an $SO(3)$ connection
rather than $SU(2)$.},
and the physical subspace
is the cotangent space of the
moduli space ${\cal{M}}(\Sigma)$
of flat $SU(2)$ connections over $\Sigma$.
Thus a physical wave function in the $ISO(3)$ CS-theory
can be regarded as a gauge-invariant function on
${\cal{M}}(\Sigma)$.
This establishes the connection between the Ponzano-Regge lattice
model and the continuum field theory \iso .

What about the $q$-analogue model?
Since the quantum $6j$-symbol gives the fusion matrix of the WZW model,
it is natural to expect that $H^{(k)}$
is related to the space ${\cal{H}}^{(k)}$
of conformal blocks of the WZW model.
When $k=1$, the structure of $H^{(k=1)}(\Sigma)$ is rather simple.
In this case, $U_q(SU(2))$ has two representations
$j=0$ or $1/2$, and an element of $C^{(k=1)}(\Sigma,t)$
may be viewed as a collection of closed
monochromatic cycles on $\Sigma$.
The constraint $Q_{t\rightarrow t}=1$ then implies that
$H^{(k=1)}(\Sigma)$ is the space of functions on
$H_1(\Sigma,{\cal{Z}}_2)$,
and it is isomorphic to ${\cal{H}}^{(k=1)}
\otimes \overline{{\cal{H}}^{(k=1)}}$.
For $k \geq 2$, we must deal with a graph with $k$ different colors,
$j=1/2,1,...,k/2$. However, if a contour in the graph is colored
by $j \geq 1$, by using a map $Q_{t \rightarrow t'}$,
it can be decomposed into a network consisting
only of contours with $j=1/2$.
We have examined the structure
of $H^{(k)}(\Sigma)$ for lower genus $\Sigma$
and found that
they are isomorphic to ${\cal{H}}^{(k)}
\otimes \overline{{\cal{H}}^{(k)}}$. Detailed of this procedure
and its extension to higher genera will be discussed elsewhere.

This suggests that the partition function $Z_k$
of the Turaev-Viro model
is equal to the absolute-value-square of the partition function
of the $SU(2)$ CS-theory when $M$ is orientable.
Recently we were informed that this had indeed been proven by
Turaev in a rather different approach\foot{We thank
T.Kohno for informing this to us and T.Takata for sending us
a copy of a
hand-written manuscript [\Turaevtwo ] by Turaev.}.
The partition
function $Z_k$ of the $q$-analogue lattice model
should then be expressed as
$$
\eqalign{ Z_k & = \left|
    \int [ dA ] \exp \left( i { k \over 4 \pi} \int ( A dA
   + {2 \over 3} A^3 ) \right) \right|^2 \cr
    & = \int [ dA, dB ] \exp \left( i{k\over 4\pi}
          \int (A dA + {2 \over 3} A^3)
                               - (B dB + {2\over 3} B^3)\right) \cr
   & = \int [ de , d\omega ] \exp \left(
    i \int e \wedge R + {\lambda_k \over 3} e \wedge e \wedge e \right)
\cr
   &~~~~~~~~~~~~~~~~~~~~ \omega = {1 \over 2}(A + B)~,~~
                     e = {k \over 8 \pi}(A-B) ~ , \cr}
\eqn\doubling
$$
where the ``cosmological constant'' $\lambda_k$ is equal to
$(4\pi / k)^2$. Due to the cosmological term, a classical
solution to $R + \lambda_k e \wedge e =0$ would be
a three-dimensional sphere of
radius $k/4\pi$. It is intriguing note that the maximum length
$k/2$ of a geodesic on the sphere
coincides with the maximum value of $j$ for $U_q(SU(2))$.
In the limit of $k \rightarrow \infty$,
the cosmological constant vanishes and the above equation
reduces to \iso .

To regard \iso\ or \doubling\ as an Euclidean functional integral for
quantum gravity, the factor $i$ in front of the action is disturbing.
One might try to eliminate it by rotating the contour of the
$e$-integral, but the resulting functional integral would then be
divergent. The problem is that the sign of $\int e\wedge e \wedge e$
is indefinite unlike $\int \sqrt{g}$, which is clearly positive
definite,
and one does not know how to define an Euclidean functional integral
in the first order formalism of $e$ and $\omega$.
The volume form could be made positive definite by integrating
over $e$ with fixed orientation only
($e \rightarrow -e$ flips an orientation
in three dimensions), but it is not clear whether the resulting
theory is renormalizable.

This issue would be addressed by studying a Lorentzian version
of the lattice model based on
infinite dimensional unitary representations of $SO(2,1)$.
A work in this direction is now in progress.

It is straightforward, at least in the limit of $k \rightarrow \infty$,
to extend the above analysis to another compact Lie group $G$.
In general there are more than one way to contract Wilson-line
operators at a vertex. This means that, in such a lattice model,
we put colors on
faces of tetrahedra in $M$ as well as on edges.
This model should be equivalent to what is called
the BF-theory\refmark{\bftheory}\foot{In three dimensions,
the $ISO(3)$ CS-theory may be regarded as the $SU(2)$ BF-theory.}
for the group $G$.

\vskip 3mm
\noindent
{\bf Acknowledgments}

The authors would like to thank
R.Sorkin for discussion on the model of Ponzano and Regge, and
A.Shapere for bringing the paper by
Turaev and Viro into their attention. They are grateful to
T.Kohno and T.Takata for information on the work by
Turaev.

\refout
\bye
\end